\documentclass[twocolumn, amsmath,amssymb,aps,pra,showpacs]{revtex4-1}

\usepackage{graphicx}
\usepackage{dcolumn}
\usepackage{bm}
\usepackage{braket}
\usepackage{mathrsfs}

\begin{document}

\title{Experimental observation of carrier-envelope phase effects by multicycle pulses.}

\author{Pankaj K. Jha,$^{1,2,*}$ Yuri V. Rostovtsev,$^{3}$ Hebin Li,$^{1,\dagger}$ Vladimir A. Sautenkov,$^{1,4}$ and Marlan O. Scully$^{1,2}$}

\affiliation{$^{1}$Institute for Quantum Science and Engineering and Department of Physics and Astronomy, Texas A\&M University, College Station,Texas 77843, USA\\
$^{2}$Mechanical and Aerospace Engineering and the Princeton Institute for the Science and Technology of Materials, Princeton University, Princeton, NJ 08544, USA\\
$^{3}$Department of Physics, University of North Texas, Denton, Texas 76203, USA\\
$^{4}$P.N.Lebedev Physical Institute of the Russian Academy of Sciences, Moscow 119991, Russia}
\pacs{32.80.Rm 42.50.-p 32.80.-t 37.10.Jk}

\begin{abstract}
\noindent We present an experimental and theoretical study of carrier-envelope phase (CEP) effects on the population transfer between two bound atomic states interacting with pulses consisting of many cycles. Using intense  radio-frequency pulse with Rabi frequency of the order of the atomic transition frequency, we investigated the influence of CEP on the control of phase dependent multi-photon transitions between the Zeeman sub-levels of the ground state of $^{87}$Rb. Our scheme has no limitation on the duration of the pulses. Extending the CEP control to longer pulses creates interesting possibilities to generate pulses with accuracy that is better then the period of optical oscillations.
\end{abstract}

\maketitle

\section{Introduction}

As is well-known, the electric field of a laser pulse given by 
\begin{equation}
E(t) = {\cal E}_0 f(t) \cos(\nu t + \phi) 
\end{equation}
can be characterized by its amplitude ${\cal E}_0$, its carrier envelope $f(t)$, its frequency $\nu$, and its carrier-envelope phase (CEP) $\phi$. The CEP is the most difficult parameter to control and even to measure. Recently, a lot of research has been devoted to the CEP. Namely, the CEP strongly affects many processes involving ultrashort few-cycle pulses~\cite{brabec00rmp}. In particular, CEP effects on high-harmonic generation \cite{boham98prl}, strong-field photoionization \cite{paulus01nature}, the dissociation of HD$^+$ and H$_2^+$ \cite{rudnev04prl2}, the electron dynamics in a strong magnetic field \cite{baluska02nature}, the population inversion during a quantum transition \cite{r9}, and the external- and internal- photo-effect currents \cite{r10,frontierprl04} have been demonstrated by few-cycle pulses. 

For longer laser pulses, the influence of the CEP becomes smaller (very often it is beyond the experimental abilities to be measured). So the important question is what is the maximal duration of laser pulses that can still have the CEP effects? It is a fundamental question, but also it brings new interesting possibilities to measure and control parameters of laser pulses and applications. A stabilized and adjustable CEP is important for applications such as optical frequency combs \cite{diddams2000} and quantum control in various media \cite{yin1992}. Several techniques have been developed to control the CEP of femtosecond pulses \cite{jones2000,baluska02nature}. A crucial step in attaining this control is measuring the CEP to provide feedback to the laser system. Promising approaches for short pulses use, for instance, photoionization \cite{verhoef06ol} and quantum interference in semiconductors \cite{frontierprl04}. 

For longer pulses, on the other hand, there are no such methods. Recently, a method has been presented for the measurement of the absolute CEP of a high-power, many-cycle driving pulse, by measuring the variation of the XUV spectrum~\cite{tzallas10prl} by applying the interferometric polarization gating technique to such pulses~\cite{NP}. We stress here that extending the CEP control to longer pulses creates interesting possibilities to generate pulses with accuracy that is better than the period of optical oscillation. First, it allows researchers to improve laser systems that generate laser pulses with better reproducibility and accuracy and better controlled. 
Second, it provides an additional handle to control the process of collisions. Femtosecond pulses are shorter than the time duration of collisions and cannot be used to study collisions under the action of electromagnetic fields; meanwhile the current approach of extending the duration of the pulses with measureable or controllable CEP allows researchers to extend the coherent control to a new level when they are able to study molecular collisions or electron collisions in nanostructures under the action of strong electromagnetic fields with known CEP. Electromagnetically induced magnetochiral anisotropy in a resonant medium demonstrated in \cite{sau05prl} can be enhanced by the control of the CEP of optical radiation in the laser induced chemical reactions~\cite{chem-lasers}. 

In this paper, we report the CEP effects in the population transfer between two bound atomic states interacting with pulses consisting many cycles in contrast with few-cycle pulses~\cite{hli10prl}. For our experiment, we use intense radio-frequency (RF) pulses interacting with the magnetic Zeeman sub-levels of Rubidium (Rb) atoms. We have found that, for long pulses consisting two carrier frequencies, the CEP of the pulse strongly affects that transfer. It is worth noting here that our scheme has no limitation on the duration of pulses.

The significance of our experiment is that it provides the insight of CEP effect in a new regime. The experiment is the first, to our knowledge, to observe the CEP effect on a transition between two \emph{bound} atomic states with such \emph{long} pulses. Our experiment provides a unique system serving as an experimental model for studying ultrashort optical pulses. The obtained results may be easily extended to optical experiments.

The paper is organized as follows. In section II, we briefly discuss the experimental setup and the procedure to determine the population transfer due to RF excitation. We present our experimental results in Figs. \ref{Fig4},\ref{Fig5},\ref{Fig6}. In section III using a simple two-level model, we explain the phase dependence of the main results presented in Fig. \ref{Fig6}. In section IV we present discussion on extending the CEP control to longer pulses. We have added an appendix with an explicit calculation of the probability amplitudes for the one and multi-photon excitation.
\begin{figure}[t]
\includegraphics[height=5.0cm,width=0.48\textwidth,angle=0]{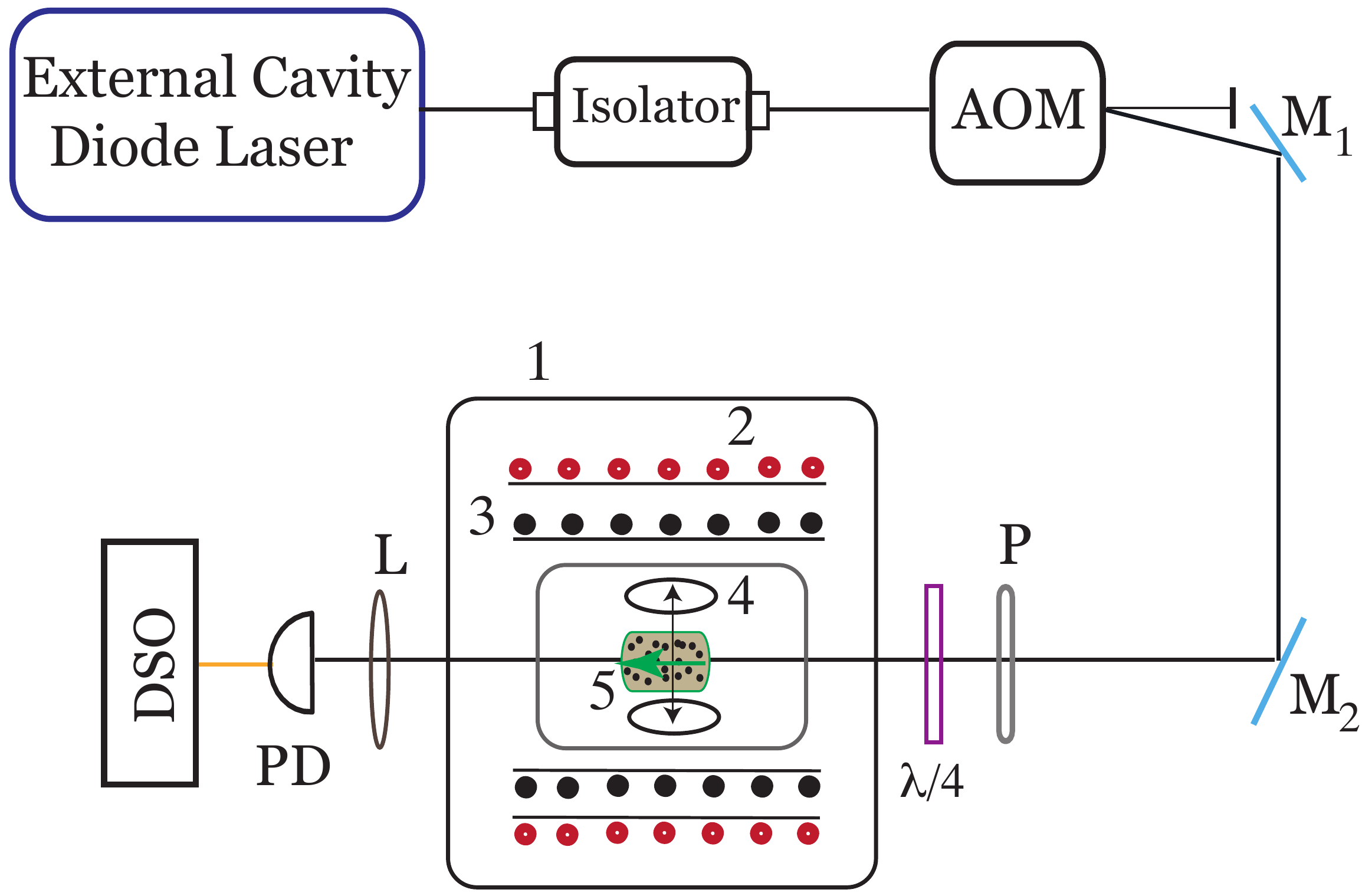}    
 \caption{(Color Online) Experimental setup. ECDL-External cavity diode laser; AOM- Acousto-optic modulator; P- Polarizer, PD-Photodiode; L-Lens, the oven is assembled with 1. copper tube; 2. non-magnetic heater on a magnetic shield; 3. solenoid; 4. pair of Helmholtz coils; 5. Rb cell.
 \label{Fig1}
 }
\end{figure}

\section{Experiment} 
In this section we will discuss the experimental aspect of our paper. We discuss the setup and the procedure to measure the population transfer due to RF excitation, taking into account the dephasing factor $\eta$. In subsection B, we present our experimental results which includes the non-linear behavior of the multi-photon excitation peak \textcircled{3} [see Fig. \ref{Fig4} (a)]. Effect of the CEP of the carrier-frequency components on the population transfer due to multi-photon excitation is shown in Fig. \ref{Fig6}.
\subsection{Setup and Population transfer}The experimental setup is shown in Fig.~\ref{Fig1}. An external cavity diode laser was tuned to the $\text{D}_{1}$ resonance line of $^{87}$Rb atoms at $| 5^{2}S_{1/2}; F=1\rangle \leftrightarrow | 5^{2}P_{1/2}; F=1\rangle$ transition. A 2.5 cm long cell containing $^{87}$Rb (and 5 torr of Neon) is located in an oven. The cell is heated in order to reach an atomic density of the order of $10^{11}$ cm$^{-3}$. A longitudinal static magnetic field is applied along the laser beam to control the splitting of the Zeeman sub-levels of the ground state $| 5^{2}S_{1/2}; F=1,m_{F}=-1,0,1\rangle$.  A pair of Helmholtz coils produces a transverse bichromatic rf field  with two central frequencies at $\nu_{1}$ and $\nu_{2}$ \cite{P}. In our experiment we tuned the longitudinal magnetic field to control the Zeeman splitting while keeping the carrier frequencies intact. A function generator was programmed to provide multi-cycle bichromatic pulses with controllable parameters, such as the pulse duration, CEPs and the amplitudes of the two carrier frequencies.

\begin{figure}[t]
    \includegraphics[height=3.5cm,width=0.48\textwidth,angle=0]{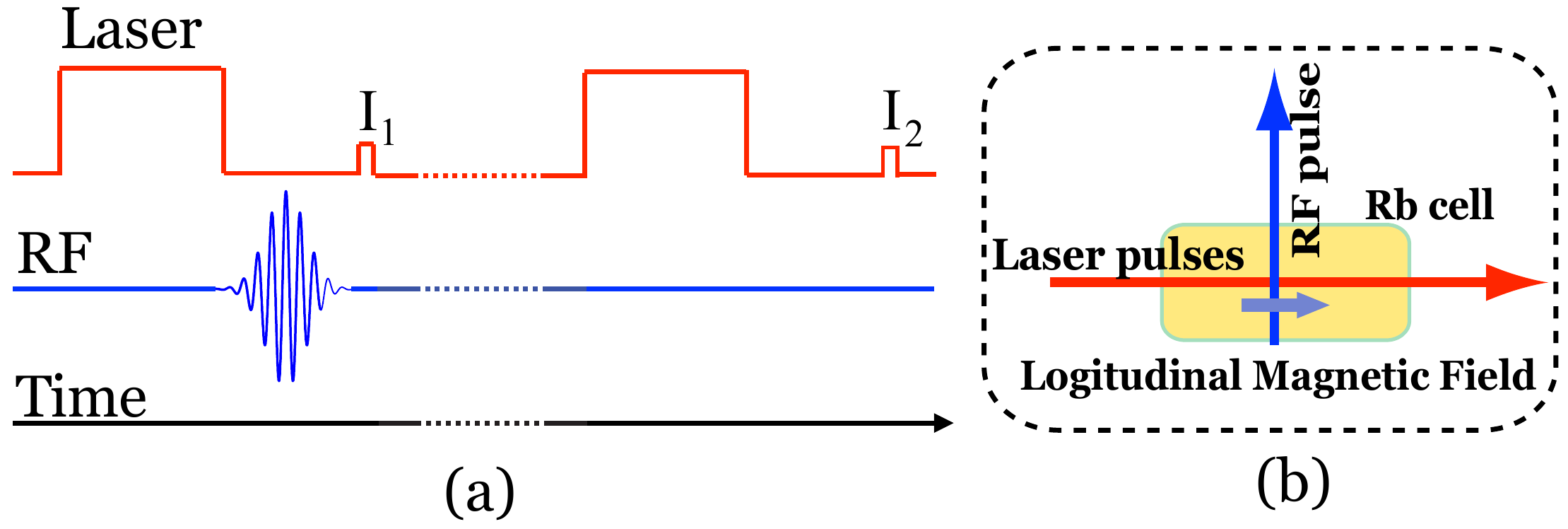}
  \caption{(Color Online) (a) Time sequence of the laser and the RF pulses to determine the population transfer due to RF excitation. (b) Configuration of the laser and rf pulses along with the longitudinal magnetic field with respect to the Rb cell. 
  \label{Fig2}
  }
\end{figure}
\begin{figure}[b]
    \includegraphics[height=3.2cm,width=0.48\textwidth,angle=0]{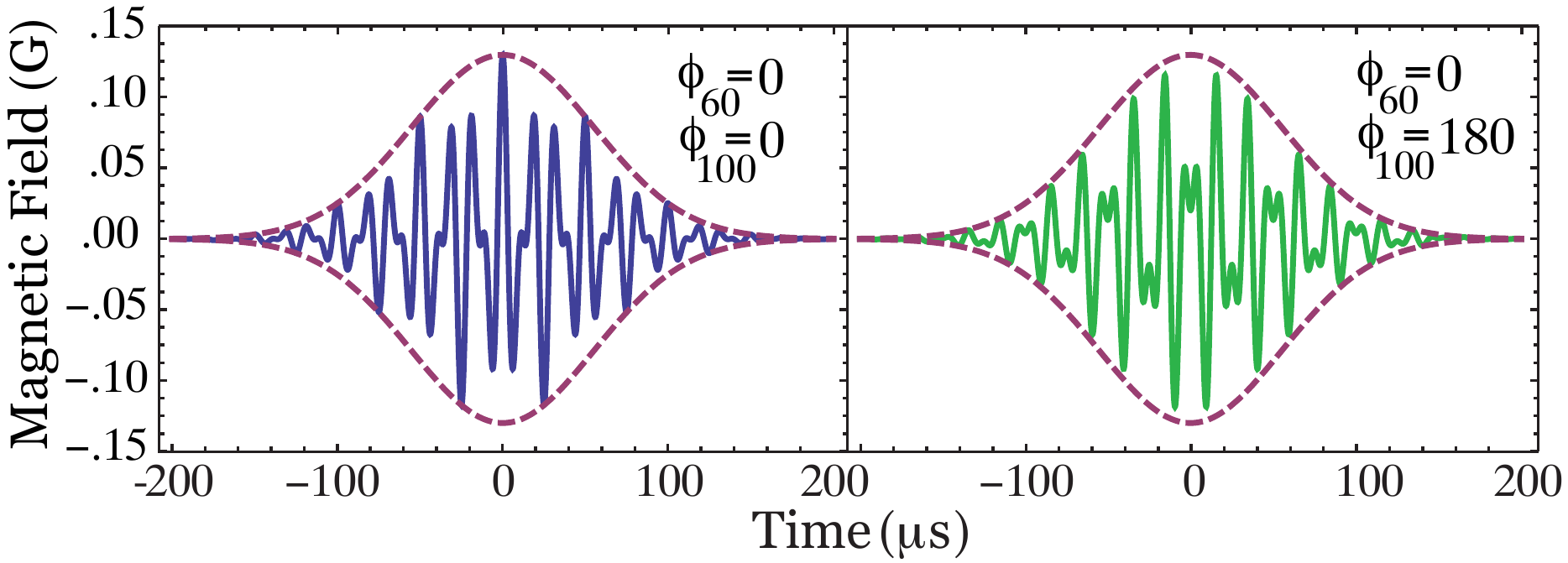}
  \caption{(Color Online) CEP-shaped bichromatic pulses with spectral components of 60 kHz and 100kHz. FWHM for both the pulse is 130 $\mu s $ with gaussian envelope. Unit of the magnetic field is Gauss.
  \label{Fig3}
  }
\end{figure}
\begin{figure*}[htb]
    \includegraphics[height=6.3cm,width=0.65\textwidth,angle=0]{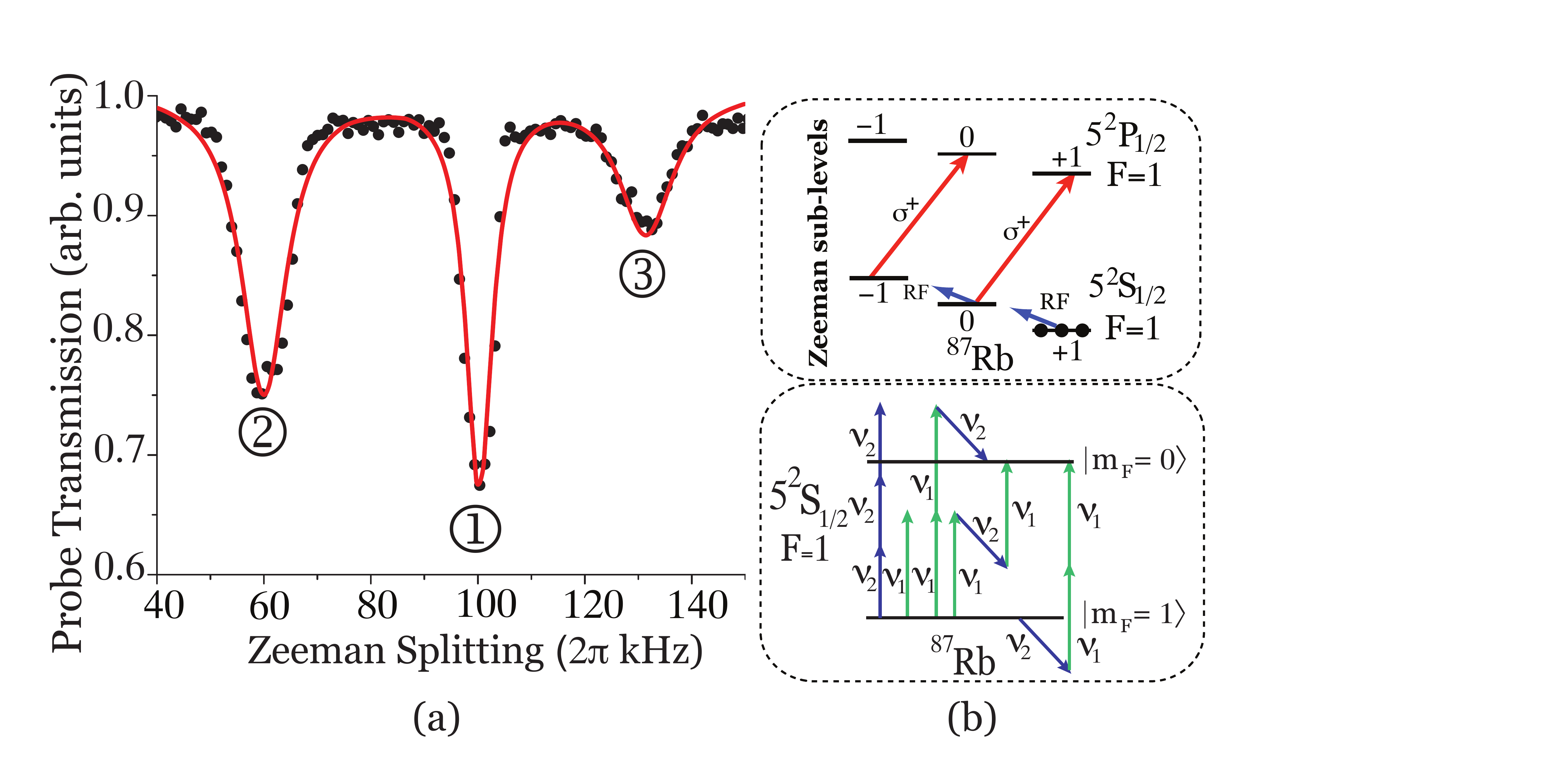}
  \caption{(Color Online) (a) Optical probe transmission profile for the one-photon [peaks \textcircled{1} and \textcircled{2}] and three-photon [peak \textcircled{3}] transition under the bichromatic rf field excitation. (b) Upper block: Energy level scheme of $^{87}$Rb; Lower block:  Resonant and non-resonant pathways contributing to three-photon peak.
  \label{Fig4}
  }
\end{figure*}
To determine the population transfer due to the rf excitation, the experiment is performed with a sequence of laser pulses with a rf pulse followed by a sequence of laser pulses without rf pulse. For the transmitted probe pulse intensity  is given by $I_{1}=I_{0}\eta e^{N\sigma LP_{a}}$, where $I_{0}$ is the probe pulse input intensity, $\eta$ is the factor due to dephasing, $N$ is the atomic density, $\sigma$ is the absorption cross-section, $L$ is the cell length and $P_{a}$ is the population of the upper levels due to RF excitation. For the second sequence , in which there is no RF excitation, the transmitted probe pulse intensity is given by $I_{2}=I_{0}\eta$. Therefore, the population due to rf excitation is given by the quantity $-ln(I_{1}/I_{2})=N\sigma L P_{a}$.
 
The energy level scheme of $^{87}$Rb and the configuration of the optical and  RF pulses is shown in Fig.~\ref{Fig2}. The ground state of $^{87}$Rb has three Zeeman sub-levels; a right-circularly polarized (RCP) laser pulse optically pumps the system and drives the atoms to the sub-level $| 5^{2}S_{1/2}; F=1,m_{F}=1\rangle$. This is followed by the bichromatic rf pulse, which excites the atoms to the sub-levels $| 5^{2}S_{1/2}; F=1,m_{F}=-1,0\rangle$ whose population is subsequently determined by measuring the transmission of a following weak RCP optical probe pulse. The rf pulse is delayed by 165 $\mu s$ with respect to the optical-pumping laser pulse and has a duration of 130 $\mu s$ (FWHM). In Fig.~\ref{Fig3} we have plotted two such CEP-shaped bichromatic pulses, with spectral components of 60 kHz and 100kHz, used in our experiment. The transmitted intensity of the probe pulse, delayed by 330 $\mu $s with respect to the optical-pumping pulse, is monitored by a fast photodiode.

\begin{figure}[b]
    \includegraphics[height=4.5cm,width=0.34\textwidth,angle=0]{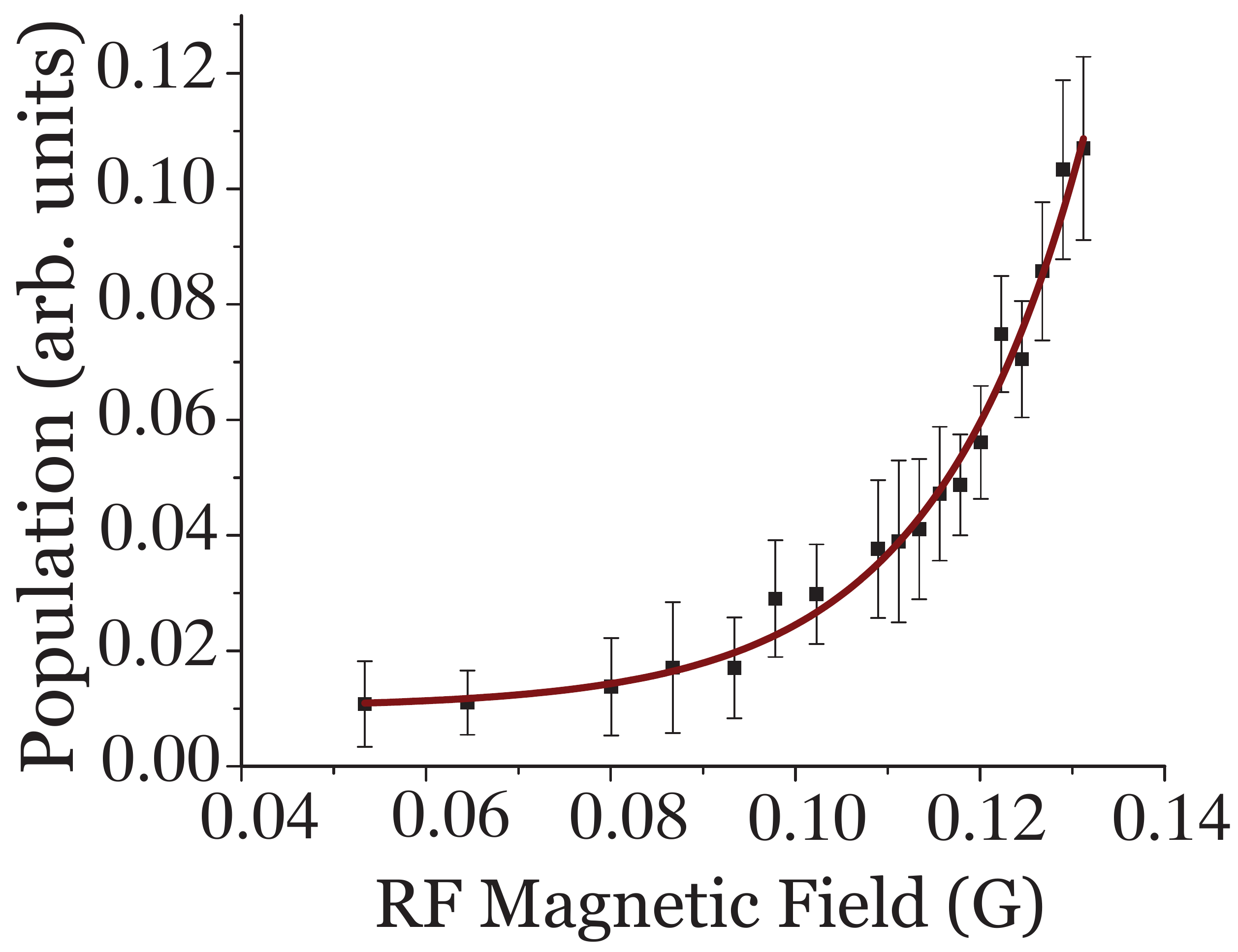}
  \caption{(Color Online) Non-linear dependence of multi-photon excitation on the traverse magnetic field. Unit of the magnetic field is Gauss.
  \label{Fig5}
  }
\end{figure}
\subsection{Experimental Results.}
Single and multi-photon (resonant and non-resonant) excitation under bichromatic rf field interaction with $^{87}$Rb are shown in Fig.~\ref{Fig4}. Peaks \textcircled{1} and \textcircled{2} in the probe transmission profile are single photon absorption peaks at frequencies $\omega_{1}$=100kHz and $\omega_{2}$=60kHz respectively. Peak \textcircled{3} emerges due to different possible excitations between the initial and the final states [see Fig.~\ref{Fig4} (b) lower block]. 
Resonant multi-photon excitation which corresponds to peak \textcircled{3} at $\omega$=140kHz in Fig.~\ref{Fig4}, is shifted to about $\omega$=130kHz \cite{ramsey49pr}. The rf field is very strong, so non-resonant one- and three-photon transition should be taken into account [see Appendix]. These non-resonant contributions interfere with resonant three-photon transitions and the excited population depends on the phases of fields with frequencies $\nu_1$ and $\nu_2$. To study this peak we first investigated the dependence of population transfer as a function of the applied transverse magnetic field strength. Fig. 5 shows the non-linear behavior of the process, in which the multi-photon excitation is negligible for weak transverse magnetic field and starts to grow non-linearly with the increase in the amplitude of the driving RF pulse.

\begin{figure*}[t]                  
  \includegraphics[height=5.5cm,width=0.8\textwidth,angle=0]{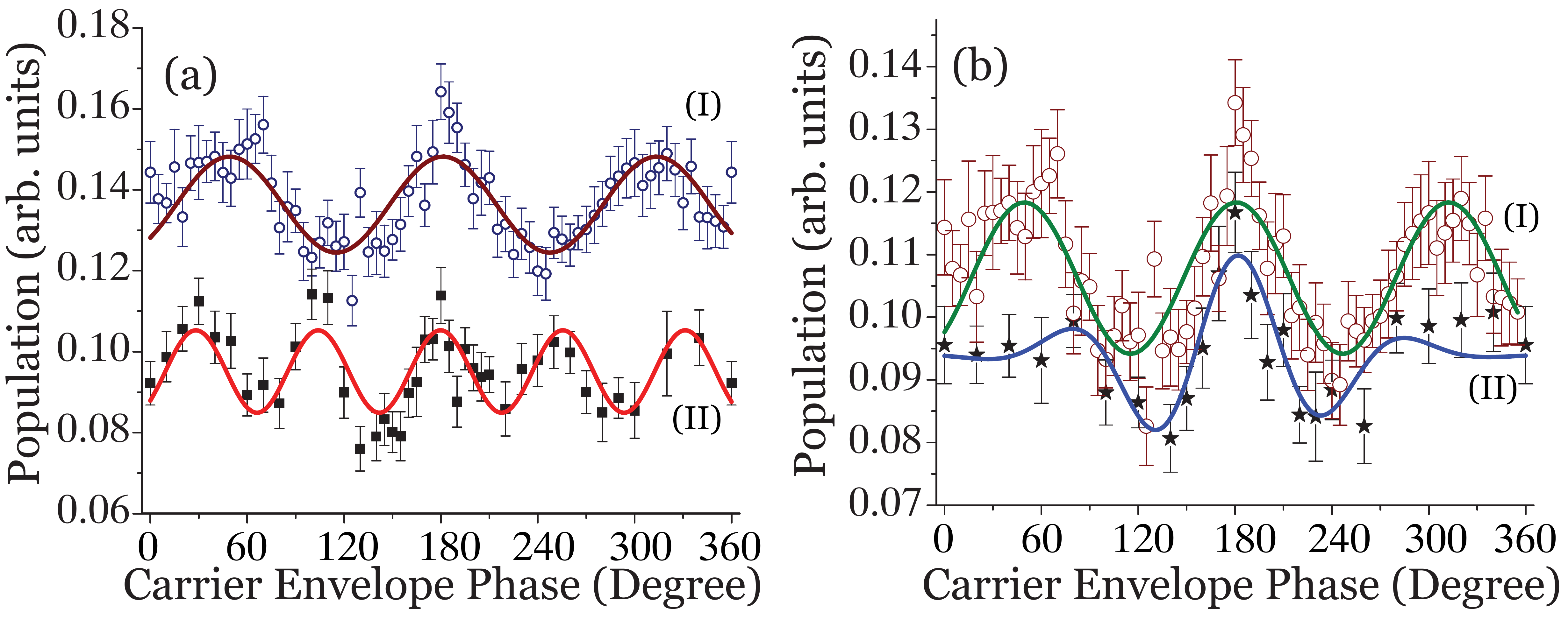}
  \caption{(Color Online) Oscillatory nature of the population transfer by changing the phase of one carrier frequency while keeping the other at zero for the bichromatic rf Pulse. (a) (I) Changing the phase $\phi_{100\text{kHz}}$ and $\phi_{60\text{kHz}}$=0 (II) Changing the phase $\phi_{60\text{kHz}}$ and $\phi_{100\text{kHz}}$=0. (b) Effect of the pulse duration T (FWHM) on the population transfer. (I) T=130$\mu s$, (II) T=100$\mu s$. Here we changed the phase $\phi_{100\text{kHz}}$ while keeping $\phi_{60\text{kHz}}=0$
  \label{Fig6}
  }
\end{figure*}

The main results of the experiment are shown in Fig.~\ref{Fig6} where we have 
plotted the population ($\sigma NLP_{a}$) as a function of carrier-envelope 
phase of one of the two spectral components of the bichromatic field while 
keeping the other phase component at zero. Fig.~\ref{Fig6}(a)(II) shows 
the oscillatory behavior when the phase of $\phi_{60 \text{kHz}}$ is changed 
while keeping $\phi_{100\text{kHz}}=0$. Similar effect is observed vice-versa 
which is shown in Fig.~\ref{Fig6}(a)(I). Ratio of the frequency of oscillations
for the two cases, when the phase is changed from $0 \rightarrow 2\pi$, is  
$O_{r}=0.578 \pm 0.035$ which is equal to $\nu_{2}/\nu_{1}$. Fig.~\ref{Fig6}(b) 
shows the effect of pulse duration (i.e number of cycles) on the population 
transfer where we have plotted the population transferred for two set of 
pulse width $T$(full width at half maximum, FWHM). Here (I) $T$=130$\mu s$, 
(II) $T$=100$\mu s$. In either case we changed the phase of 
$\phi_{100\text{kHz}}$ while keeping $\phi_{60\text{kHz}}=0$. 
In Fig.~\ref{Fig6} (a) we have shifted the curve (I) vertically, for the sake of clarity 
and distinguish the variations in the two curve (I) \& (II) clearly.

\section{Theory}

Let us now move to the theoretical aspect of the results obtained here. 
The goal of theoretical consideration presented here is to gain 
physical insights that helps to understand the CEP effects for such long pulses
that have envelop containing up to fifteen periods of oscillations, as well as
the limitations imposed on the length of pulses.  
The Hamiltonian for an atomic state with $F=1$ in a magnetic field $B=(B_{x},B_{y},B_{z})$ is given by
\begin{equation}\label{e1}
\mathscr{H}=-g\mu_{0}\left( {\begin{array}{ccc}
 B_{z} & \frac{B_{x}+iB_{y}}{\sqrt{2}} & 0  \\
 \frac{B_{x}-iB_{y}}{\sqrt{2}} & 0 & \frac{B_{x}+iB_{y}}{\sqrt{2}}   \\
 0 & \frac{B_{x}-iB_{y}}{\sqrt{2}} & -B_{z}  \\
 \end{array} } \right),
\end{equation}
where $g=-1/2$ is the Lande factor for this Rb state, $\mu_{0}$ is the Bohr magneton,  $B_{z}=B_{0}$ is the static magnetic field that is chosen in the direction of the z-axis;  $B_{x}$ and $B_{y}$ are the transverse components driven by a function generator. The linearly-polarized bichromatic magnetic field is given as,
\begin{equation}\label{e2}
B_{x}(t)=e^{-\alpha ^{2}t^{2}}\{B_{1}\text{cos}(\nu_{1}t+\phi_{1})+B_{2}\text{cos}(\nu_{2}t+\phi_{2})\},
\end{equation}
where $\alpha = (2\sqrt{\text{ln}2})/T$ and $T$ is the FWHM duration of the pulse and $B_{y}=0$. For the magnetic dipole transition, the relaxation due to atomic motion is the most important. The density matrix equations is given by
\begin{equation}\label{e3}
\dot{\rho}=-\frac{i}{\hbar}[\mathscr{H},\rho]-\Gamma (\rho-\rho_{0}),
\end{equation}
where $\mathscr{H}$ is given by Eq.(\ref{e1}), $\Gamma$ quantifies the relaxation process due to atomic motion and $\rho_{0}$ is the thermal equilibrium density matrix of the atoms in the cell without the optical and RF fields. For simple explanation we will consider only two levels coupled by the bichromatic field and neglect any type of relaxation. The Rabi frequency is given by
\begin{equation}\label{e4}
\Omega(t)=e^{-\alpha ^{2}t^{2}}\{\Omega_{1}\text{cos}(\nu_{1}t+\phi_{1})+\Omega_{2}\text{cos}(\nu_{2}t+\phi_{2})\},
\end{equation}
where $\Omega_{(1,2)}=g\mu_{0}B_{(1,2)}/\sqrt{2}\hbar$. The equation of motions for the probability amplitudes $C_{a}$ and $C_{b}$ are given by~\cite{A3,A4}
\begin{subequations}\label{e5}
\begin{align}
\dot{C}_{a}&=i\Omega(t)e^{i\omega t}C_{b}, \label{second}\\
\dot{C}_{b}&=i\Omega^{*}(t)e^{-i\omega t}C_{a}.
\end{align}
\end{subequations}
Let us consider the perturbative approach $C_{b}(t) \cong 1$. We look for a solution of the form $C_{a}=C_{a}^{(1)}+C_{a}^{(3)}$. The excited population is the result of interference of resonant three-photon excitation and non-resonant one-photon with frequency $\nu_1$ and three-photon $\nu_2$ where the detunings are 30~kHz and 
50~kHz correspondingly [see inset of Fig.~\ref{Fig4}(a)]. The probability amplitude can be written as
\begin{equation}
C_a = A_{1}(\nu_1) e^{-i\phi_1} + A_{3}(\nu_2)e^{-i3\phi_2} + A_3(2\nu_1-\nu_2)e^{-i(2\phi_1 - \phi_2)}
\label{eq1-exp1}
\end{equation}
that gives the same dependences on the phases of bichromatic field as shown in Fig.~\ref{Fig6}. Here, in a weak field approximation, 
\begin{equation}
A_1(\nu_1)=i\left(\frac{\sqrt{\pi}}{2\alpha}\right)\Omega_{1}e^{-[(\omega-\nu_{1})/2\alpha]^{2}}e^{-i\phi_{1}},
\label{eq2-exp1}
\end{equation}
is the probability amplitude of non-resonant excitation due to one-photon transition, 
\begin{equation}
A_3(\nu_2)=-i\left[\frac{\sqrt{\pi}\Omega^{3}_{2}}
{16\sqrt{3}\alpha\nu_{2}(\omega-\nu_{2})
}\right]e^{-\left[(\omega -3\nu_{2})^{2}/12\alpha^{2}\right]-3i\phi_{2}} ,
\label{eq3-exp1}
\end{equation}
is due to non-resonant three-photon excitation, and 
\begin{equation}
\begin{split}
A_3(2\nu_1 - \nu_2)& =-i\left(\frac{\sqrt{\pi}\Omega^{2}_{1}\Omega_{2}}{8\sqrt{3}\alpha}\right)\left[\frac{1}{2\nu_{1}(\omega-\nu_{1})}+\right.\\
&\left. \frac{1}{(\nu_{1}-\nu_{2})(\omega-\nu_{1})}+\frac{1}{(\nu_{1}-\nu_{2})(\omega+\nu_{2})}\right]\\
&\times e^{-\left[(\omega -2\nu_1+\nu_{2})^{2}/12\alpha^{2}\right]-2i\phi_{1} + i\phi_2}
\end{split}
\label{eq4-exp1}
\end{equation}
is due to resonant three-photon excitation. Here the first terms corresponds 
to Hyper-Raman type process, the second term corresponds to Doppleron type 
process as shown in the lower block of Fig. {\ref{Fig4} (b)}. In Appendix A 
we have shown the relative strength of the three processes with 
the experimental parameters. 

As is clearly seen from Eq.(\ref{eq1-exp1}), the CEP effect occurs 
due to the interference of the terms that have different dependence on 
the field phases. The condition for the better visibility of the interference 
is related to the amplitudes and frequencies of fields. It is better to have 
amplitude be the same to have high visibility, on the other hand, 
if only one term dominates the CEP effect disappears. It is very 
interesting to note here that the CEP effects do not depend explicitly 
on the duration of pulses but only on the field amplitudes and 
their frequencies. 

\section{conclusion}

 We use intense RF pulses interacting with the magnetic Zeeman sub-levels of Rubidium (Rb) atoms, we have experimentally and theoretically shown the CEP effects in the population transfer between two bound atomic states interacting with pulses consisting of many cycles (up to 15 cycles) of the field. It opens several exciting applications and interesting possibilities that can be easily transfer to optical range and enhance current and create new set of tools to control CEP of laser pulses.
  
These tools allow researchers to improve laser systems that generate laser pulses with better reproducibility and accuracy and better controlled. Also the tools provide an additional handle to control the process of collisions, and the current approach of extending the duration of the pulses with measurable or controllable CEP allows researchers to extend the coherent control to a new level where they are able to study molecular collisions or electron collisions in nano-structures under the action of strong electromagnetic fields with known CEP. In particularly, the obtained results can be applied to  control of chemical reactions~\cite{chem-lasers}. 

\section{Acknowledgment}
We thank  L.V.~Keldysh, O.~Kocharovskaya,  T.~Siebert and M.S.~Zubairy  for useful discussions and gratefully acknowledge the support from the NSF Grant EEC-0540832 (MIRTHE ERC), Office of Naval Research (N00014-09-1-0888 and N00014-08-1-0948), Robert A. Welch Foundation (Award A-1261)), Herman F. Heep and Minnie Belle Heep Texas A$\&$M University Endowed Fund held/administered by the Texas A$\&$M Foundation and Y.V.R. gratefully acknowledges the support from the UNT Research Initiation Grant and the summer fellowship UNT program. 

\noindent$^*$Email: pkjha@physics.tamu.edu\\
$\dagger$Current Address: JILA University of Colorado, 440 UCB Boulder, CO 80309-0440, USA

\appendix
\section{Single and Multi-Photon Excitation Probability Amplitudes}
The wave function of a two-level atom can be written in the form
\begin{equation}\label{P1}
|\psi(t)\rangle=C_{a}(t)e^{-i\omega_{a}t}|a\rangle + C_{b}(t)e^{-i\omega_{b}t})|b\rangle,
\end{equation}
where $C_{a}$ and $C_{b}$ are the probability amplitudes of finding the atom in the states $|a\rangle$ and $|b\rangle$, respectively. The equation of motions for $C_{a}$ and $C_{b}$ are given by,
\begin{equation}\label{P2}
\dot{C}_{a}(t)=i\Omega(t)e^{i\omega t}C_{b}(t)
\end{equation}
\begin{equation}\label{P3}
\dot{C}_{b}(t)=i\Omega^{*}(t)e^{-i\omega t}C_{a}(t).
\end{equation}
Integrating Eq.(\ref{P2}) we obtain
\begin{equation}\label{P4}
C_{a}(t)=i\int_{-\infty}^{t}\Omega(t')e^{i\omega t'}C_{b}(t')dt'
\end{equation}
In the limit $t\rightarrow \infty$ Eq.(\ref{P4}) gives,
\begin{equation}\label{P5}
C_{a}(\infty)=i\int_{-\infty}^{\infty}\Omega(t')e^{i\omega t'}C_{b}(t')dt'
\end{equation}
\begin{widetext}
Substituting Eq.(\ref{P4}) in Eq.(\ref{P3}) and using the initial condition $C_{b}(0)=1$ we get,
\begin{equation}\label{P6}
C_{b}(t')=1-\int_{-\infty}^{t'}\left[\Omega^{*}(t'')e^{-i\omega t''}\left(\int_{-\infty}^{t''}\Omega(t''')e^{i\omega t'''}C_{a}(t''')dt'''\right)dt''\right]
\end{equation}
Plugging back Eq.(\ref{P6}) in Eq.(\ref{P4}), we get
\begin{equation}\label{P7}
C_{a}(t)=i\int_{-\infty}^{t}\Omega(t')e^{i\omega t'}\left\{1-\int_{-\infty}^{t'}\left[\Omega^{*}(t'')e^{-i\omega t''}\left(\int_{-\infty}^{t''}\Omega(t''')e^{i\omega t'''}C_{a}(t''')dt'''\right)dt''\right]\right\}dt'
\end{equation}
Thus from Eq.(\ref{P7}) we get,
\begin{equation}\label{P8}
C_{a}(\infty)=i\int_{-\infty}^{\infty}\Omega(t')e^{i\omega t'}\left\{1-\int_{-\infty}^{t'}\left[\Omega^{*}(t'')e^{-i\omega t''}\left(\int_{-\infty}^{t''}\Omega(t''')e^{i\omega t'''}C_{a}(t''')dt'''\right)dt''\right]\right\}dt'
\end{equation}
In the perturbation theory $C_{b}(t) \cong 1$, we are looking for a solution of the form $C_{a}(\infty)=C_{a}^{(1)}(\infty)+C_{a}^{(3)}(\infty)$, where the first term $C_{a}^{(1)}(\infty)$ is given by
\begin{equation}\label{P9}
C_{a}^{(1)}(\infty)=i\int_{-\infty}^{\infty}\Omega(t')e^{i\omega t'}dt'
\end{equation}
The second term can be found as
\begin{equation}\label{P10}
\begin{split}
C_{a}^{(3)}(\infty)=-i\int_{-\infty}^{\infty}\left\{\Omega(t')e^{i\omega t'}\int_{-\infty}^{t'}\left[\Omega^{*}(t'')e^{-i\omega t''}\int_{-\infty}^{t''}\Omega(t''')e^{i\omega t'''}dt'''\right]dt''\right\}dt'
\end{split}
\end{equation}
Let us consider that the Rabi frequency $\Omega(t)$ is given as 
\begin{equation}\label{P11}
\Omega(t)=e^{-\alpha ^{2}t^{2}}\{\Omega_{1}\text{cos}(\nu_{1}t+\phi_{1})+\Omega_{2}\text{cos}(\nu_{2}t+\phi_{2})\},
\end{equation}
\subsection{Single Photon Processes}
(i) Absorption of one-photon of frequency $\nu_{1}$. The transition probability amplitude is given as 
\begin{equation}\label{P12}
C_{a,(\nu_{1})}^{(1)}(\infty)=i\left(\frac{\sqrt{\pi}}{2\alpha}\right)\Omega_{1}e^{-[(\omega-\nu_{1})/2\alpha]^{2}}e^{-i\phi_{1}}
\end{equation}
Similarly we can find $C_{a,(\nu_{2})}^{(1)}(\infty)$ using the substitution $\Omega_{1}\rightarrow \Omega_{2}, \nu_{1}\rightarrow \nu_{2}$ and $\phi_{1}\rightarrow \phi_{2}$.
\subsection{Multi-Photon Processes}
(ii) Absorption of three-photon of frequency $\nu_{2}$. The transition probability amplitude is given as 
\begin{equation}\label{P13}
C_{a,(\nu_{2},\nu_{2},\nu_{2})}^{(3)}(\infty)=-i\left[\frac{\sqrt{\pi}}{16\sqrt{3}\alpha\nu_{2}(\omega-\nu_{2})}\right]\Omega^{3}_{2}e^{-(1/3)[(\omega-3\nu_{2})/2\alpha]^{2}}e^{-3i\phi_{2}}
\end{equation}

 (iii) Absorption of two-photon of frequency $\nu_{1}$ and emission of one-photon of frequency $\nu_{2}$ in the order:
 
 (iii.a) $\nu_{1}\rightarrow\nu_{1}\rightarrow\nu_{2}$. The transition probability amplitude is given as 
\begin{equation}\label{P14}
C_{a,(\nu_{1},\nu_{1},\nu_{2})}^{(3)}(\infty)=-i\left[\frac{\sqrt{\pi}}{16\sqrt{3}\alpha\nu_{1}(\omega-\nu_{1})}\right]\Omega^{2}_{1}\Omega_{2}e^{-(1/3)[(2\nu_{1}-\nu_{2}-\omega)/2\alpha]^{2}}e^{-i[2\phi_{1}-\phi_{2}]}
\end{equation}

(iii.b) $\nu_{1}\rightarrow\nu_{2}\rightarrow\nu_{1}$. The transition probability amplitude is given as 
\begin{equation}\label{P15}
C_{a,(\nu_{1},\nu_{2},\nu_{1})}^{(3)}(\infty)=-i\left[\frac{\sqrt{\pi}}{8\sqrt{3}\alpha(\nu_{1}-\nu_{2})(\omega-\nu_{1})}\right]\Omega^{2}_{1}\Omega_{2}e^{-(1/3)[(2\nu_{1}-\nu_{2}-\omega)/2\alpha]^{2}}e^{-i[2\phi_{1}-\phi_{2}]}
\end{equation}

(iii.c) $\nu_{2}\rightarrow\nu_{1}\rightarrow\nu_{1}$. The transition probability amplitude is given as 
\begin{equation}\label{P16}
C_{a,(\nu_{2},\nu_{1},\nu_{1})}^{(3)}(\infty)=-i\left[\frac{\sqrt{\pi}}{8\sqrt{3}\alpha(\nu_{1}-\nu_{2})(\nu_{2}+\omega)}\right]\Omega^{2}_{1}\Omega_{2}e^{-(1/3)[(2\nu_{1}-\nu_{2}-\omega)/2\alpha]^{2}}e^{-i[2\phi_{1}-\phi_{2}]}
\end{equation}
\end{widetext}
The resonant three-photon excitation we are investigating are given by (iii.a), (iii.b) and (iii.c). Let us find the ratio of the amplitudes $R_{\alpha}$ for the processes (iii.a) and (iii.b) defined as 
\begin{equation}\label{P17}
R_{\alpha}=\frac{\big|C_{a,(\nu_{1},\nu_{1},\nu_{2})}^{(3)}(\infty)\big|}{\big|C_{a,(\nu_{1},\nu_{2},\nu_{1})}^{(3)}(\infty)\big|}
\end{equation}
gives
\begin{equation}\label{P18}
R_{\alpha}=\frac{\nu_{1}-\nu_{2}}{2\nu_{1}}
\end{equation}
This ratio $R_{\alpha} \rightarrow 0$ in the limit $\nu_{1}\rightarrow \nu_{2}$ i.e Doppleron type process given by Eq.(\ref{P15}) dominates over the hyper-Raman type process given by Eq.(\ref{P14}) and other resonant and non-resonant processes. Similarly the ratio of the amplitudes $R_{\beta}$ for the processes (iii.c) and (iii.b) defined as 
\begin{equation}\label{P19}
R_{\beta}=\frac{|C_{a,(\nu_{2},\nu_{1},\nu_{1})}^{(3)}(\infty)|}{|C_{a,(\nu_{1},\nu_{2},\nu_{1})}^{(3)}(\infty)|}
\end{equation}
gives
\begin{equation}\label{P20}
R_{\beta}=\frac{\omega-\nu_{1}}{\omega+\nu_{2}}
\end{equation}
In this case smaller the one photon detuning $\omega-\nu_{1}$, greater will be the probability of the Doppleron type process. The ratio of the amplitudes $R_{\gamma}$ for the processes (ii) and (i) defined as 
\begin{equation}\label{P21}
R_{\gamma}=\frac{|C_{a,(\nu_{2},\nu_{2},\nu_{2})}^{(3)}(\infty)|}{|C_{a,(\nu_{1})}^{(1)}(\infty)|}
\end{equation}
gives
\begin{equation}\label{P22}
R_{\gamma}=\frac{\Omega^{3}_{2}e^{[(\omega-3\nu_{2})^{2}/6\alpha^{2}]}}{8\sqrt{3}\Omega_{1}\nu_{2}(\omega-\nu_{2})}
\end{equation}
Ratio of the amplitudes $R_{\delta}$ for the processes (i) and (iii.b) defined as 
\begin{equation}\label{P23}
R_{\delta}=\frac{|C_{a,(\nu_{1})}^{(1)}(\infty)|}{|C_{a,(\nu_{1},\nu_{2},\nu_{1})}^{(3)}(\infty)|}
\end{equation}
gives
\begin{equation}\label{P24}
R_{\delta}=\frac{4\sqrt{3}(\nu_{1}-\nu_{2})(\omega-\nu_{1})e^{[-(\omega-\nu_{1})^{2}/4\alpha^{2}]}}{\Omega_{1}\Omega_{2}}
\end{equation}
For small $\alpha$, this ratio is very small and we can neglect the contribution of the non-resonant one-photon excitation with respect to the resonant three-photon excitation to a good approximation. But for large $\alpha$ i.e small pulse duration we should be careful. Let us consider $\Omega_{2} \approx 0.3\nu_{1}, \Omega_{1} \approx 0.4\nu_{1}, \nu_{2}=0.6\nu_{1},\omega=1.4\nu_{1}$ and $\alpha \approx 0.128\nu_{1}$. Using this parameters we obtain $R_{\delta} \approx 0.8$
We obtain $R_{\delta} \approx 0.8$, thus absorption of one-photon of $\nu_{1}$ followed by emission of one-photon of $\nu_{2}$ followed by absorption of one-photon of $\nu_{1}$ is comparable to one-photon absorption of $\nu_{1}$. Thus we can see the contribution of off-resonant one-photon absorption to Peak $\textcircled{3}$ is not negligible.  


\begin{thebibliography}{99}
\bibitem{P} Here we use the convention that all frequencies are circular frequencies so that $\hbar\nu$ (not $h\nu$) is the photon energy and the  atomic transition frequency is $\omega=\omega_{a}-\omega_{b}$. We define detuning as $\Delta=\omega-\nu$.
\end{thebibliography}
\end{document}